# Unraveling capillary interaction and viscoelastic response in atomic force microscopy of hydrated collagen fibrils

Manuel R. Uhlig*[a,b] and Robert Magerle*[a,c]

[a] Fakultät für Naturwissenschaften, Technische Universität Chemnitz, D-09107 Chemnitz Germany
[b] Present address: Instituto de Ciencia de Materiales de Madrid (CSIC), c/ Sor Juana Inés de la Cruz 3, 28049 Madrid, Spain. E-mail: manuel.uhlig@csic.es
[c] E-mail: robert.magerle@physik.tu-chemnitz.de

## Abstract

The mechanical properties of collagen fibrils depend on the amount and the distribution of water molecules within the fibrils. Here, we use atomic force microscopy (AFM) to study the effect of hydration on the viscoelastic properties of reconstituted type I collagen fibrils in air with controlled relative humidity. With the same AFM tip, we investigate the same area of a collagen fibril with two different force spectroscopy methods: force-distance (FD) and amplitude-phase-distance (APD) measurements. This allows us to separate the contributions of the fibril's viscoelastic response and the capillary force to the tip–sample interaction. A water bridge forms between the tip apex and the surface, causing an attractive capillary force, which is the main contribution to the energy dissipated from the tip to the specimen in dynamic AFM. The force hysteresis in the FD measurements and the tip indentation of only 2 nm in the APD measurements show that the hydrated collagen fibril is a viscoelastic solid. The mechanical properties of the gap regions and the overlap regions in the fibril's D-band pattern differ only in the top 2 nm but not in the fibril's bulk. We attribute this to the reduced number of intermolecular crosslinks in the reconstituted collagen fibril. The presented methodology allows the mechanical surface properties of hydrated collagenous tissues and biomaterials to be studied with unprecedented detail on the nanometer scale.

## Introduction

The collagens are the most abundant protein family in vertebrates and a major constituent of connective tissues like bone, cartilage, tendon, and skin.[1] Type I collagen forms 300-nm-long triple helical protein complexes (tropocollagens) which self-assemble into collagen fibrils with a characteristic stripe pattern. This D-band pattern results from a staggered arrangement of adjacent tropocollagen molecules caused by a periodic sequence of amino acids along the backbone of type I tropocollagen.[2] The pattern repeats every 67 nm and consists of a gap region and an overlap region. Water molecules are an integral partof the molecular structure of collagens, and the mechanical properties and the structure of collagen ils and collagen-based tissues depend sensitively on the amount and the distribution of water in the interprotein space. For an overview, see the book edited by Fratzl[1] and the reviews of Kadler *et al.*,[3] Gauteri *et al.*[4] and Sherman *et al.*[5]

The distribution of water in collagen tissues and its influence on the collagen structure has been addressed by different approaches, such as infrared spectroscopy,[6] nuclear magnetic resonance,[7] differential scanning calorimetry,[8] X-ray diffraction[9], and computer simulations.[4,10] With a combination of experimental and computational methods, Masic *et al.* revealed the conformational changes in collagen fibrils that occur upon water removal.[11] With molecular dynamics simulations, Streeter *et al.*[10] found two species of water molecules in type I collagen fibrils: Free water molecules fill the space between the tropocollagen molecules, whereas bound water molecules serve as bridges for hydrogen bonds between the tropocollagen molecules.

On the nanometer scale, methods based on atomic force microscopy[12] (AFM) provide a unique possibility for exploring the morphology and the mechanical properties of hydrated collagen fibrils.[13–24] Studies on individual collagen fibrils form the basis for an in-depth understanding of AFM data on structurally more complex collagenous tissues.[14,16,20,25–28] AFM-based methods have been used for imaging dry collagen fibrils[13,29,30] and for measuring their local elastic modulus and viscoelastic properties.[15,18,22,24,31–33] With nanoindentation and AFM-based tip indentation, the fibril's mechanical response perpendicular to the fibril's axis is measured, whereas with microtensile testing experiments, the mechanical response along the fibril's axis is probed. Neugirg *et al.*[34] reviewed these methods.

The effect of hydration on the fibril's elastic modulus, viscoelastic properties, and shape has been investigated when the fibrils were in aqueous solution[18,19,21] and in humid air.[22] The local elastic modulus and viscoelastic properties in the D-band gap regions and overlap regions have also been investigated in the dry state,[32,33] in humid air,[22] and in water.[21] With AFM-based nanoindentation, Grant *et al.*[18,19] and Minary-Jolandan *et al.*[32] measured the elastic modulus of type I collagen fibrils extracted from bovine Achilles tendon. Grant *et al.* found that the elastic modulus decreases from 1.9 ± 0.5 GPa in the dry state to 1.2 ± 0.1 MPa in buffer solution.[18] Minary-Jolandan *et al.* found that the elastic modulus of a dry fibril is significantly larger in overlap regions (2.2 GPa) than in gap regions (1.2 GPa).[32] Baldwin *et al.* mapped the nanomechanical properties of native collagen fibrils from rat tail tendon submersed in water and found the elastic modulus in the overlap regions to be 25% larger than in the gap regions.[21] Furthermore, they found variations in the mechanical properties along the fibril's axis on larger length scales. To study hydrated collagen fibrils, researchers have immersed collagen fibrils in buffer solution[18,19] and water[21] during AFM measurements. Moreover, the water content of a collagenous sample can be controlled by the relative humidity of the surrounding air,[22,35,11] which allows researchers to investigate the effects of increasing hydration on the structure and the mechanical properties of collagen fibrils. With multi-set point intermittent contact (MUSIC) mode AFM, Spitzner *et al.* investigated the nanoscale swelling behavior of reconstituted type I collagen fibrils in humid air. After increasing the relative humidity (RH), they observed a larger water uptake in the gap regions compared to the overlap regions, accompanied by a larger damping of the tip oscillation in the gap regions.[22] This provides direct evidence for a larger amount of free water in gap regions than in overlap regions.

Probing a sample's viscoelastic response in humid air with AFM requires a detailed understanding of the tip–sample interaction. The sketches in Fig. 1 illustrate the interaction between a hydrated collagen fibril and the AFM tip apex during the measurement of a force-distance (FD) curve. During an FD measurement, the force between the tip and the sample is measured while the cantilever approaches the sample and then retracts. The resulting FD curve yields direct information about the tip–sample contact mechanism and contains information about the sample's properties, such as the adhesion and elastic modulus.[36] In humid air, the situation is more complex, since a water layer adsorbs on the surface of hydrophilic materials,[37–39] such as the $SiO_x$-covered surface of the AFM tip and probably also on the collagen fibril, since collagen is also hydrophilic.[40] As the tip approaches the hydrated collagen fibril, at the position $d_{On}$, a water bridge forms between the collagen fibril and the AFM tip apex, resulting in an attractive capillary force (right panel). This mechanism has been identified for different hydrophilic surfaces (glass, $SiO_x$-covered Si, and mica)[39,41–43], and we interpret the FD data measured on hydrated collagen fibrils along this line. The tip reaches the collagen surface at $d_S$ (center panel), where the maximal capillary force $F_C$ acts in the approach curve. As it approaches further, the tip indents into the collagen fibril (left panel), which is accompanied by a repulsive force. As the tip retracts, the repulsive force decreases with hysteresis due to the collagen fibril's viscoelastic response. When the geometric contact between the tip and the fibril is lost, a water bridge is pulled off the surface until the distance $d_{Off}$ is reached, whereupon the water bridge breaks.

During intermittent contact (IC) mode AFM (also known as tapping mode AFM), the capillary interaction occurs during each oscillation cycle at the inflection point of the oscillating tip, when the tip taps onto the sample's surface. If the sample's surface is compliant, the tip will also make an indentation. At an oscillation frequency of 300 kHz, the contact time with the sample is only ~1 µs in contrast to ~100 ms in the case of an FD-curve measurement. In IC-mode AFM, the equivalent to an FD curve is the measurement of the amplitude and the phase of the oscillating cantilever while the tip–sample distance is reduced.[44,45] The resulting amplitude-phase-distance (APD) curves allow for the separation of the elastic and dissipative contributions to the effective tip–sample interaction, making it possible to draw conclusions regarding the sample's viscoelastic properties.[46–48] Furthermore, the APD curves yield the z-axis position of the unperturbed surface as well as the tip indentation.[22,44,49–51] On hydrated collagen fibrils, the tip indentation reaches up to 3.5 nm.[22] Spitzner et al.[45] demonstrated how height and phase images can be reconstructed for multiple set points from one data set of APD curves measured within a specified surface area. Thus, this data acquisition and data analysis mode is called multi-set point intermittent contact (MUSIC) mode AFM. The reconstructed MUSIC-mode AFM images are free from feedback loop artifacts and allow for quantitative and depth-resolved imaging of soft polymeric surfaces with unprecedented spatial resolution and detail.[22,44,45,52–54]

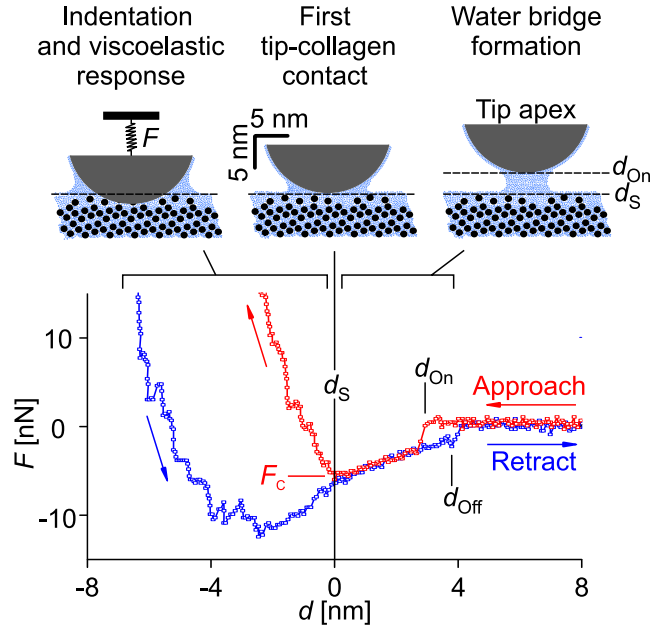

**Fig. 1.** Force–distance curve measured on a collagen fibril at 71% RH. Red and blue arrows indicate the approach curve and retraction curve, respectively. The sketches illustrate the tip–sample interaction. The tip apex is shown at the position $d_{On}$, where a water bridge between the tip and the sample forms (right); at the position $d_S$, where the tip reaches the collagen surface (center); and when it indents in the collagen fibril (left). Black dots and blue dots correspond to tropocollagen and water molecules, respectively. The view is along the fibril's axis.

Here, we use this methodology to study the role of water with regard to the nanomechanical properties of a collagen fibril as we gradually increase its water content by increasing the relative humidity (RH) of the surrounding air. We measure FD[36] and APD curves,[44,45] which provide information about the fibril's nanomechanical properties on different depth and time scales (Table 1). All the measurements are performed within the same area of the same collagen fibril to avoid natural variations in the fibril's properties, caused by the fibril's size,[55] the preparation method,[17,56] and the type of tissue.[55] Furthermore, we use the same tip for all measurements to enable a direct comparison of the results obtained from the FD and APD measurements.

We study reconstituted type I collagen fibrils, which serve as a model system with a well-defined composition and sample preparation protocol. The absence of substances such as non-collagenous proteins, fatty acids, and proteoglycans makes it possible to probe the viscoelastic response of a fibril comprising only tropocollagen and water molecules. From multicomponent polymer mixtures, it is known that one component can accumulate at the surface due to surface segregation,[60,61] forming a wetting layer which differs in chemical composition and mechanical properties from the bulk. In analogy, we expect that on a hydrated collagen fibril, water may accumulate in a surface layer, changing the viscoelastic properties of the top surface layer compared to the fibril's bulk. Our combination of FD and APD measurements allows us to quantify the contributions of the capillary interaction and the fibril's viscoelastic response to the tip–sample interaction in FD and APD measurements.

**Table 1.** AFM-based force spectroscopy methods and the corresponding accessible phenomena, indentation range, and contact times.

| Method | Accessible phenomena | Indentation range[a] | Contact time[a] |
|---|---|---|---|
| Force vs. distance[36] | Water bridge[42]<br>Adhesion force[57]<br>Elastic modulus[58] | 0–15 nm | ~100 ms |
| Amplitude and phase vs. distance[44,45] | Energy dissipation[47]<br>Effective tip–sample interaction[48] | 0–2 nm | ~1 μs[59] |

[a] Typical values in our experiment.

# Experimental section

## Sample preparation

We investigated acid-extracted, purified type I collagen from bovine calf hide (Collagen A1, MATRIX BioScience GmbH, Mörlenbach, Germany). It was purchased as an aqueous solution with a concentration of 0.9–1.5 mg/ml and a pH of 2.0–2.4 as specified by the manufacturer. The sample preparation was similar to that in previous work.[18,19] Briefly, 2 μl of collagen solution were injected in a drop of buffer solution (30 μl, L-glycine/KCl, pH 9.2) placed on a freshly cleaved mica disc to initiate the self-assembly of collagen fibrils. After 20 min, the fibrils were transferred to a polished Si substrate with native $SiO_x$ layer. Prior to fibril deposition, the Si substrate was cleaned for 20 min in a 1:1 solution of acetone/toluene followed by cleaning with a $CO_2$ jet. During the entire fibril formation and fibril deposition procedure, the substrate temperature was held constant at 37 °C using a heated sample holder (JPK ECCell, JPK Instruments AG, Berlin, Germany). After preparation, the sample was allowed to dry for 23 h at ambient conditions.

## Humidity control

All measurements were performed under controlled humidity (within an error of 1% RH) in a lab-built setup, similar to that used in previous work.[22] The humidity was regulated by the air flow through a water-filled washing bottle. The RH and the temperature was measured with a sensor (SHT75, Sensirion AG, Staefa, Switzerland) next to the sample. The first measurements took place at 44% RH; then the RH was increased stepwise to 71%, 80%, and 84% RH. At each step, the system was allowed to equilibrate for 1 h before AFM measurements were performed.

Finally, the air flow was switched off and, after 20 h, AFM measurements were performed at 40% RH. During the entire experiment, the temperature was within the range of 23.4 ± 0.8 °C.

**AFM measurements**

A NanoWizard II AFM (JPK Instruments AG, Berlin, Germany) and one single silicon cantilever (Pointptobe NCH, NanoWorld AG, Neuchâtel, Switzerland) were used for all AFM measurements except for the data measured on a clean Si substrate. The typical tip radius was < 8 nm, as specified by the manufacturer. At ambient conditions (44% RH), the cantilever had a resonance frequency $f_0$ = 298.944 kHz, a quality factor $Q$ = 383, and a spring constant $k_1$ = 19.68 N/m as determined with Sader's method.[62] During the entire experiment, the variation in these values was less than 3%. The data measured on a clean Si substrate were measured using a different cantilever of the same type with a spring constant $k_2$ = 35.54 N/m as determined with Sader's method.[62]

First, a clean silicon substrate was mounted in the AFM as a reference system, and then 49 APD and 49 FD curves were measured. Next the collagen sample was mounted, imaged with IC-mode AFM, and an appropriate fibril was selected. At each step in RH, APD and FD curves were measured on rectangular arrays, each with 60 × 45 positions separated by 8 nm, resulting in an image size of 480 x 360 nm². Each pointwise measurement was followed by a conventional IC-mode scan with an amplitude set point $A/A_0$ = 0.9 to check for drift and plastic deformations. Thus, within small positional variations, every map shows the same spot of one individual fibril. During IC-mode AFM with a set point $A/A_0$ = 0.9, the tip indentation $\tilde{z}_B$ into the fibril increases with increasing RH by 0.5 nm, whereas on the collagen lawn, the tip indentation remains constant. This results in a systematic underestimation of $h_p$. Therefore, the degree of swelling is a lower bound to the real value. The effect, however, is of the same order as the surface roughness.

**Force distance mapping**

For acquiring FD curves, the cantilever approached the sample with a speed of 0.5 μm/s until a force of 100 nN was reached and then retracted with the same speed. The tip–sample distance $d$ was determined from the piezoelectric actuator position $z$, taking the cantilever bending into account.[36] The approach curves were analyzed according to the modified Hertz model[63] including net adhesion, similar to the Derjaguin-Müller-Toporov (DMT) theory of contact mechanics.[64] Assuming a parabolic tip which is much stiffer than the sample, Sneddon[65] derived the following relation between the force $F$ and the tip–sample distance $d$:

$$F(d) = \frac{4}{3}\frac{E}{(1-v^2)}\sqrt{R}(-d)^{3/2} - F_C \quad (1)$$

Here, $E$ is the elastic modulus and $v$ the Poisson's ratio of the sample. $R$ is the curvature of the tip apex. For fitting eqn (1) to the FD data, we assumed $R$ = 10 nm, set $v$ = 0.5 as in the work of Grant et al.,[18,19] and identified the net adhesion force $F_C$ with the minimal force during approach (see Fig. 1). At this point, we set $d_S \equiv d = 0$. The fit range was limited to the first 5 nm of indentation. This corresponds to 17% of the fibril's height. For such a small indentation, the influence of the substrate on the fitting result of the elastic modulus $E$ is negligible.[24] For a cantilever force constant of 20 N/m and forces < 10 nN (100 nN), the lateral tip motion caused by the cantilever bending is smaller than 0.2 nm (2 nm).[66] For a sample modulus of 100 MPa, the resulting cantilever deflection error is ~5%, for 2 GPa, it is ~40%. The error resulting from the assumptions about the tip shape and the tip radius is of similar size, if not larger. For measuring the elastic modulus with higher accuracy, these effects should be accounted for.

**MUSIC-mode AFM**

The MUSIC-mode protocol is described elsewhere.[45] For obtaining the APD curves, the cantilever was excited to oscillate at $f_D = 0.9999\, f_R$ with a free amplitude $A_0 = 60$ nm. Then the cantilever approached the sample until an amplitude set point $A/A_0 = 0.5$ was reached, except for the measurement at 80% RH, where we reached only $A/A_0 = 0.78$. From the amplitude $A$ and phase shift $\varphi$, we calculate the tip indentation, $\tilde{z}_0$ and $\tilde{z}_B$, according to refs 22,49,50, the conservative contribution to the tip–sample interaction, $k_{TS}$,[48] as a function of the mean tip–sample distance, as well as the amount of energy dissipated per oscillation cycle, $E_{Dis}$,[46,47] as a function of the amplitude ratio $A/A_0$.

## Results and discussion

We first present our results on the fibril's morphology and its swelling behavior in water vapor. To this end, we use conventional IC-mode AFM. Then the results from the FD and APD measurements are presented, which yield information on the fibril's nanomechanical properties on different depth and time scales (Table 1) as well as on the contribution of the capillary interaction to the tip–sample interaction in the FD and APD measurements. Finally, we briefly report the results of the FD and APD measurements on the substrate area, both next to the fibril and on a clean Si substrate. These data taken on non-deformable surfaces serve as a reference and corroborate the interpretation of the data obtained on the deformable (viscoelastic) collagen fibril.

**Morphology and swelling behavior**

Fig. 2 shows IC-mode AFM height and phase images of collagen fibrils adsorbed on a clean silicon substrate at 44% RH at the beginning of the experiment. The fibrils form a network similar to those in previous studies.[19,22,67,68] The largest fibril shows a D-band stripe pattern in the height and the phase image. Additionally, the phase image (Fig. 2b) reveals details of the surface structure. Next to the fibrils, the substrate is covered with a fine structure, which has been previously observed in other studies.[22,67–69] We will refer to it as the collagen lawn,[22,69] because it presumably consists of non-assembled tropocollagen molecules and residual constituents of the collagen solution. The phase image further shows the intersection of two fibrils (Arrow 1 in Fig. 2b), spots on the collagen lawn with a higher phase signal (Arrow 2), and a defect in the fibril structure (Arrow 3). These features serve to locate the same position throughout the entire experiment. The yellow rectangle marks the position where all the FD and APD data were taken.

Fig. 2c shows the phase image at the end of the experiment at 40% RH. Comparing this phase image with the phase image prior to the experiment (Fig. 2b) reveals two findings: First, the image quality in the area outside the yellow frame remains unchanged. Hence, the tip shape stays the same throughout the entire experiment. Second, the collagen fibril in the area enclosed by the yellow frame (Fig. 2c) has a morphology similar to its initial morphology (Fig. 2b), with only a slight broadening, and the D-band pattern appears blurred. This shows that almost all tip-induced deformation of the collagen fibril is reversible within a timescale of 10 min. Furthermore, the defect (Arrow 3 in Fig. 2b) vanished. This indicates that the tropocollagens could rearrange in the hydrated state due to a higher molecular mobility and the repeated mechanical stimulus from the AFM tip. This is probably also the reason for the slight broadening of the fibril.

To quantify the fibril's water uptake, we investigated the fibril's swelling behavior in water vapor with increasing RH. Fig. 2d shows the collagen fibril's height profiles at 44% RH (green line) and 84% RH (blue line). The profiles were extracted from IC-mode AFM height images at the position marked by the yellow line in Fig. 2a. At 44% RH, the fibril's height is 30 nm, which we refer to as $h_0$. At 84% RH, the height has increased to 35 nm. However, the fibril's width does not change during swelling. A possible explanation is that the adhesion forces between the collagen fibril and the underlying substrate pin the three-phase contact lines (Arrows 4 and 5 in Fig. 2b). Fig. 2e shows the degree of swelling, $h_p/h_0$, as a function of the RH. With increasing RH, $h_p/h_0$ increases gradually. At 84% RH, the fibril's height is 17% larger than at 44% RH, which is in line with the results of Spitzner *et al.*[22] (indicated by the cross in Fig. 2e). The small difference is attributed to differences between the fibrils studied and to Spitzner *et al.* normalizing the relative height to the height at 28% RH. After the subsequent decrease in the RH from 84% to 40%, the observed decrease in height (indicated by the gray filled circle in Fig. 2e) shows that the swelling is reversible within 20 h.

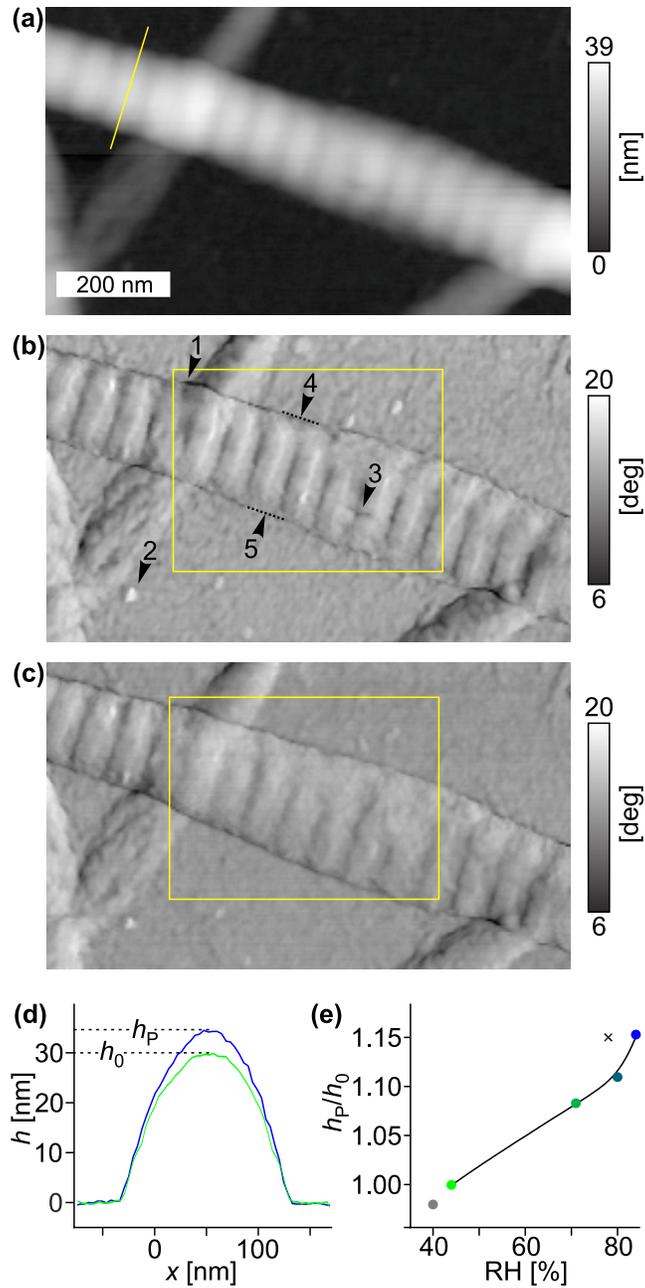

**Fig. 2.** (a-c) IC-mode AFM images of a collagen fibril deposited on a silicon substrate, taken at an amplitude set point $A/A_0 = 0.9$. (a) Height image and (b) phase image at the beginning of the experiment at 44% RH. (c) Phase image taken at the end of the experiment at 40% RH. The yellow rectangles indicate where the viscoelastic response was mapped. (d) Profiles extracted from the height images at the position marked by the yellow line in (a). The green curve and the blue curve correspond to the initial profile at 44% RH and the profile at 84% RH, respectively. (e) The degree of swelling, $h_P/h_0$, as a function of the RH as determined from the height profiles. The gray filled circle is the data point at the end of the experiment; the cross indicates the degree of swelling found by Spitzner et al.[22]. The solid line serves as a guide to the eye

**Force-distance measurements**

Fig. 1 shows a FD measurement on a collagen fibril at 71% RH. During the cantilever approach (red curve), the curve shows a sudden onset of an attractive force when the tip reaches $d_{On}$. This can be explained by the formation of a water bridge between the tip apex and the sample, which causes an attractive capillary force.[42,43] As the cantilever approaches further, the attractive capillary force increases linearly until the maximum attractive force is reached, which we refer to as $F_C$. We note that $F_C$ is the net attractive force, which also includes the van der Waals adhesion between the tip apex and the sample. The tip's downward movement shortens the water bridge, which causes an increase in the water bridge's circumference and an increase in the capillary force.[36,42] At $d = d_S$, the attractive force starts to decrease due to the onset of repulsive forces. Hence we interpret the position $d_S$ as the first contact between the tip apex and the fibril's surface. For $d < d_S$, the tip indents into the fibril. For $d < d_S$, the retraction curve (blue curve) does not coincide with the approach curve. The hysteresis depends on the cantilever's speed during the approach and the retraction from the surface. This shows that the fibril's mechanical response is viscoelastic. However, for $d_S < d < d_{On}$, the approach curve and the retraction curve match each other exactly, indicating that the water bridge can be reversibly shortened and stretched.[42] The exact match between the curve branches also shows that the accumulation of additional water due to further condensation of water vapor and the transport of liquid water during the presence of the water bridge is negligible. This is in line with the finding of Sirghi *et al.* that a water bridge is already stable after 1 ms when the AFM tip radius is small.[42] During tip retraction, the water bridge persists up to a distance $d_{Off} > d_{On}$. Hence the water bridge formation shows hysteresis accompanied by energy dissipation.[41,47] We note that the reversible formation of the water bridge can only be observed in FD curves by using a stiff cantilever that avoids the snap-to-contact that is typical for soft cantilevers.[36] The distances $d_{On}$ and $d_{Off}$ range from 3 to 8 nm (depending on the RH) and are much larger than the typical thickness of the adsorbed liquid water layers on hydrophilic materials (up to 0.3 nm at 80% RH).[37,38] Therefore, during the tip's approach, the water bridge forms before the geometrical contact between the adsorbed water layers.

Fig. 3a shows FD curves measured on a Si substrate (black) at 44% RH and on the collagen fibril's ridge (in color) at different RHs. A detail of the curve measured on the collagen fibril at 71% RH is shown in Fig. 1. Up to 80% RH, each curve shows the features which we ascribe to the interaction between the tip apex and a water bridge. This shows that capillary interaction dominates the contact mechanism for $d > d_S$ up to 80% RH. At 84% RH, the transition from the attractive interaction via the water bridge to the onset of repulsive forces (at $d = d_S$) is very smooth, which may indicate the existence of a soft collagen surface. Upon retraction, an almost constant attractive force is measured up to a distance of 30 nm, terminated by a distinct rupture. A possible explanation is the pulling out of the collagen molecules or the formation of a polymer bridge[70,71] between the tip apex and the collagen fibril. However, IC-mode AFM images taken after FD mapping do not show any changes in the fibril's surface morphology. This shows that any possible deformation of the collagen fibril due to the FD measurements relaxes within a period of approximately 10 min.

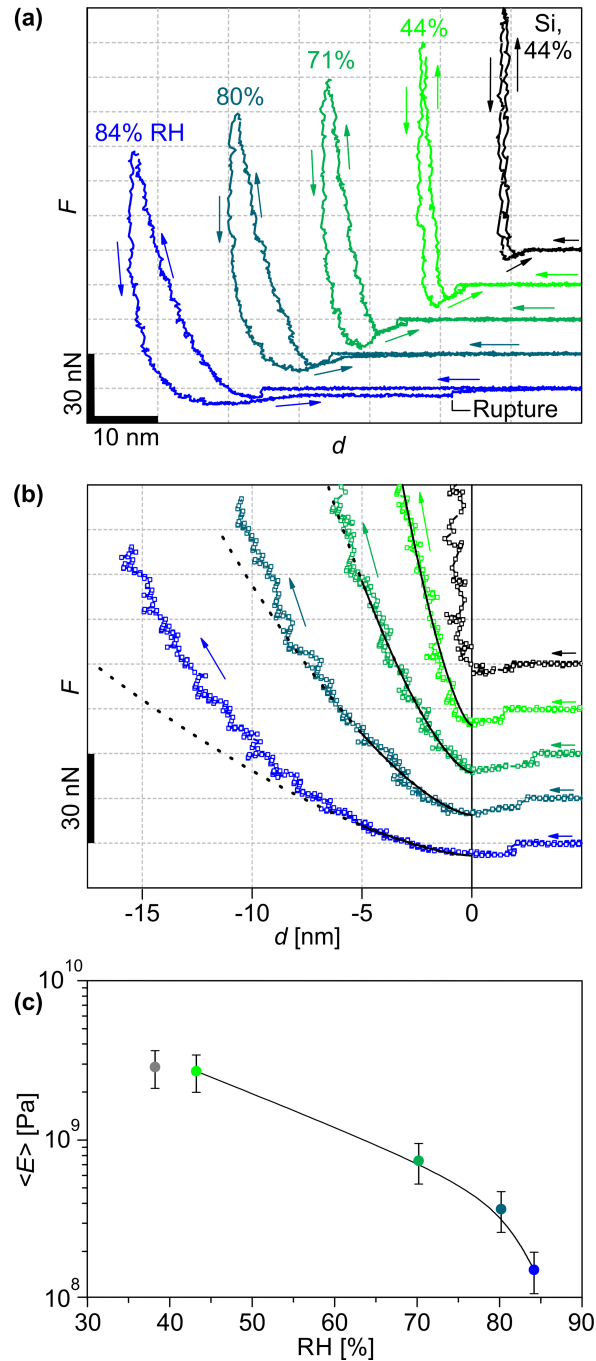

**Fig. 3.** (a) Force-distance curves measured on silicon (black) and on the collagen fibril (in color). The RH is indicated for each curve, and the arrows indicate the course of time. The curves were shifted diagonally for clarity. (b) Approach curves extracted from (a), vertically stacked for clarity and shifted such that $d_s = 0$ nm. The solid lines show fit curves for the data based on a modified Hertz model that includes adhesion. The dashed lines are the extrapolation of the fit beyond the fit range. (c) The average elastic modulus, $<E>$, obtained from the fit curves plotted as a function of the RH. The error bars represent the standard deviation. The gray filled circle is the data point at the end of the experiment. The solid line serves as a guide to the eye.

**Tip Indentation and Elastic Modulus.** Fig. 3b shows the approach curve branches of the FD curves shown in Fig. 3a. With increasing RH, the slope of the curves decreases gradually, indicating a softening of the fibril. To determine the fibril's elastic modulus, we fitted curves based on the Hertz model with adhesion additionally included [eqn (1)] to the data within the indentation range 0–5 nm. The fit curves (solid lines) describe the data well. Fig. 3c shows the average elastic modulus $<E>$ as a function of the RH, determined from all the FD curves taken on the fibril's ridge (with 100 to 250 curves for each RH value). At 44% RH, $<E>$ = 2.63 ± 0.69 GPa, which is in line with the elastic moduli of collagen fibrils in dry air (1.9 ± 0.5 GPa) observed by Grant et al.[18] With increasing RH, the elastic modulus decreases gradually. At 84% RH, $<E>$ = 0.149 ± 0.04 GPa, which is two orders of magnitude larger than 1.2 ± 0.1 MPa, the value reported by Grant et al. for collagen fibrils immersed in buffer solution.[18]

The dashed lines are the extrapolation of the fit curve beyond the fit range, which describe the data well up to 80% RH. However, at 84% RH, the fit systematically underestimates the measured force for $d < -5$ nm, where the indentation is larger than the fit range. One possible explanation is that the hydrated fibril is stiffer in the bulk than in the near-surface region, presumably due to a lower water concentration in the bulk. However, a more likely explanation is the substrate's influence, which causes an additional repulsive force for indentation depths larger than 10 nm, corresponding to 1/3 of the fibril's local height. Measurements on collagen fibrils[24,31] and on thin polymer films[58,72] show that the apparent elastic modulus determined from FD curves increases with increasing tip indentation. In our case, for RH ≥ 71%, increasing the fit range to 10 nm of indentation yields (on average) 25% larger elastic moduli (data not shown).

Fig. 4b and 4c show maps of the maximum tip indentation $d$(100 nN) and the elastic modulus $E$, which were reconstructed from the FD curves. With increasing RH, the maximum indentation measured on the fibril gradually increases, while the elastic modulus in the corresponding region gradually decreases. The difference in indentation between the gap regions and the overlap regions is very small. This was also observed by Grant et al.[18] For RH ≥ 71%, the gap and overlap regions do not differ in the elastic modulus, indicating that mechanical properties are homogenous in the indentation range of 0−5 nm. For the 10-nm fit range, we obtain the same result (not shown). The missing contrast in the elastic modulus between the gap regions and overlap regions for RH ≥ 71% is different than in native hydrated collagen fibrils.[21,32,33] We propose that this effect is due to the reduced number of crosslinks between the tropocollagen molecules in the reconstituted collagen fibril. The fibril's ridge shows smaller indentations and a higher elastic modulus than the fibril's side regions, indicating a higher stiffness of the ridge as also reported by Baldwin et al. for native collagen fibrils.[21] At the end of the experiment at 40% RH (bottom row), the indentation and elastic modulus maps show less contrast between the gap and overlap regions than at the beginning at 44% RH (top row); however, the fibril's average elastic modulus has been restored (Fig. 3c). This indicates that the changes in the fibril's mechanical properties were reversible. The reduced contrast between gap and overlap regions is probably caused by the mechanical stimulus with the AFM tip, which indented up to 15 nm deep into the collagen fibril (Fig. 3b) during the FD measurements. On the collagen lawn, the average indentation increases from 0.8 ± 0.2 nm at 44% RH to 1.2 ± 0.4 nm at 84% RH. The apparent elastic modulus is larger than 10 GPa, however, we note that the Hertz model is not suitable for describing the FD curves measured on the collagen lawn because of the negligible tip indentation.

**Capillary Interaction.** Fig. 4a shows maps of the capillary force $F_C$, corresponding to the maximum attractive force in the FD approach curves. At 44% RH, the map shows a D-band contrast, indicating that the capillary force is larger in gap regions than in overlap regions. Furthermore, the areas corresponding to the three-phase contact lines (the fibril's edges) show a slightly larger capillary force than the areas corresponding to the collagen lawn (the top-right and the bottom-left regions in Fig. 4a). A possible explanation is based on the fibril's surface corrugations, which affect the capillary forces as follows.

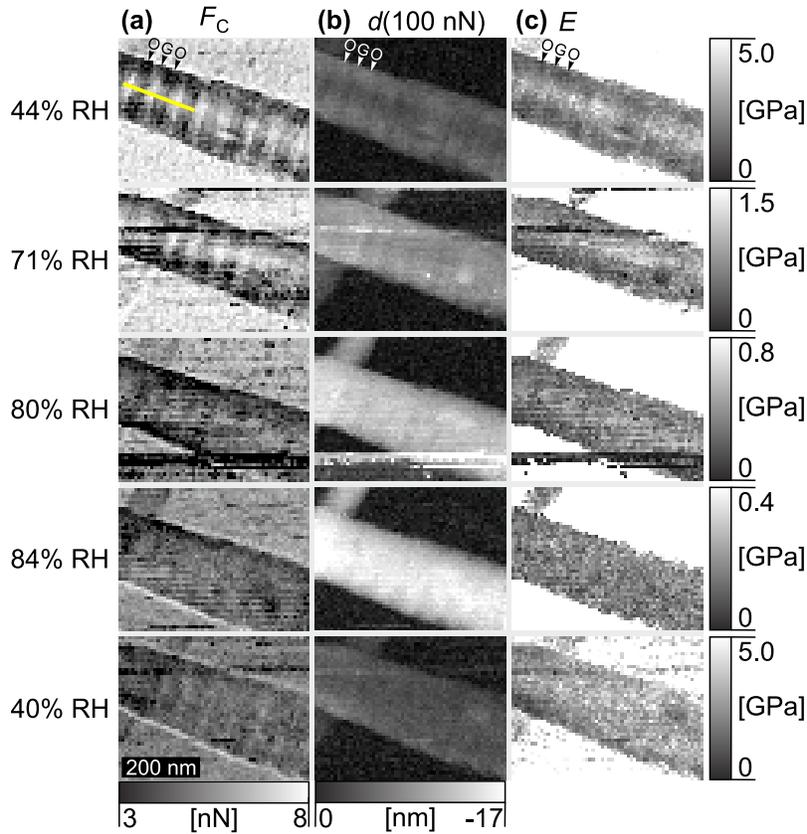

**Fig. 4.** Maps of (a) the capillary force $F_C$, (b) the maximum tip indentation during the acquisition of the force distance curves $d$(100 nN), (c) the elastic modulus $E$. The maps were measured in the area marked in Fig. 1b. For each row, the RH is indicated on the left. The yellow line marks the position of the height profile shown in Fig. 5a. The gap and overlap regions are labelled with G and O, respectively.

Fig. 5a shows the fibril's height profile at 44% RH (black) determined from the FD curves taken at the positions along the yellow line shown in Fig. 4a. The local minima and maxima in the height profile correspond to gap regions and overlap regions, respectively. The capillary force $F_C$ measured along the height profile, is shown in red. In gap regions, the average $F_C$ is 27% larger than in overlap regions. An explanation for the difference can be found in the local curvature of the fibril's surface. Fig. 5b shows a detail of the surface profile from Fig. 5a and a scale model of the tip apex, which we depict as a sphere with radius $R = 10$ nm. When the tip apex reaches the fibril's surface, the water molecules (blue dots) form a meniscus which surrounds the contact area. Sirghi et al. showed that the meniscus' radius is larger for a concave local curvature than for a convex local curvature.[73] Therefore, a meniscus formed in gap regions has a larger radius $r_{M,G}$ than a meniscus formed in overlap regions with radius $r_{M,O}$. Since the capillary force is proportional to the meniscus' radius,[36] $F_C$ is larger in gap regions than in overlap regions. In the same way, the concave geometry at the three-phase contact lines results in an increased capillary force compared to that in the adjacent areas covered with the collagen lawn.

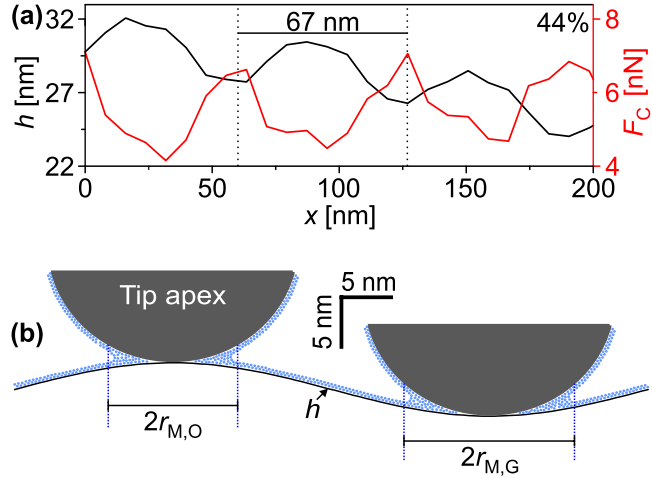

**Fig. 5.** (a) Height profile along the ridge of the collagen fibril determined from an array of force-distance curves (black line) and the corresponding capillary force $F_C$ measured during the tip approach (red). The data were measured at 44% RH, and the position of the profile is indicated in Fig. 4. (b) Effect of the height undulation on the size of the water meniscus forming between the tip and the collagen fibril (according to the method of Sirghi et al.[73]).

With increasing RH, the D-band contrast in the maps of the capillary force diminishes, whereas at the three-phase contact lines, $F_C$ remains larger than in the adjacent areas. Spitzner et al. showed that the hydration of collagen fibrils reduces the height undulation between gap regions and overlap regions and the fibril's surface becomes smooth.[22] Therefore, with increasing RH, the D-band undulation of the capillary force diminishes. In contrast, the concave geometry at the three-phase contact lines becomes even more pronounced due to swelling (Fig. 2d). Hence the capillary force measured at the three-phase contact lines increases.

**Amplitude-phase-distance measurements.**

APD measurements probe the tip–sample interaction and the fibril's viscoelastic properties in the top 2 nm, with the AFM tip vibrating at ~300 kHz with a free amplitude $A_0$ = 60 nm. Within each oscillation cycle, the tip–sample contact time is typically 1 μs.[59]

**Tip Indentation.** The tip indentation determined from APD curves depends on the tip oscillation parameters[44,51,74] and on the definition of the unperturbed surface.[22,44,49–51] Following Spitzner *et al.*, we consider the z-axis position $z_B$, where the repulsive and attractive interactions balance each other, as the unperturbed surface.[22] Fig. 6a shows the average tip indentation $<\tilde{z}_B>$ measured on the Si substrate (black) and the collagen fibril's ridge (in color) as a function of the tip–sample distance $d$ at different RHs. With decreasing tip–sample distance $d$, the average tip indentation $<\tilde{z}_B>$ increases. In our discussion on the effect of the relative humidity, we consider the average tip indentation $<\tilde{z}_B>$ measured at $A/A_0$ = 0.6, which is close to the lowest amplitude ratio $A/A_0$ reached in the APD measurements (Fig. 6b). The average tip indentation $<\tilde{z}_B>$ gradually increases from 1.5 nm at 44% RH to 2.1 nm at 84% RH, which corresponds to 1 to 2 layers of tropocollagen molecules. Spitzner *et al.* observed similar indentation values (3.1 ± 0.2 nm at $A/A_0$ = 0.6) on reconstituted collagen fibrils at 78% RH.[22] In comparison, the average indentation values $<\tilde{z}_B>$ measured on the collagen lawn with increasing RH increase from 0.3 to 0.5 nm. The corresponding plots of $<\tilde{z}_B>$ as a function of the tip–sample distance are shown in the subsection Collagen lawn and Si substrate (see below). On soft polymeric materials and polymer melts, similar tip oscillation parameters (cantilever type, tip radius, free amplitude $A_0$, and amplitude ratio $A/A_0$) lead to indentations of up to 20 nm[44,51,74], which is much larger than the values observed for hydrated collagen fibrils. Therefore, we conclude that the collagen fibril is still a solid material at 84% RH, which is in line with the results obtained from the FD data. The small indentation values show that the tip indentation during APD measurements is limited to the fibril's topmost surface layer. Hence we consider the maximum tip indentation value $<\tilde{z}_B>$ as a measure for the thickness of a compliant surface layer and conclude that its thickness increases gradually with increasing RH.

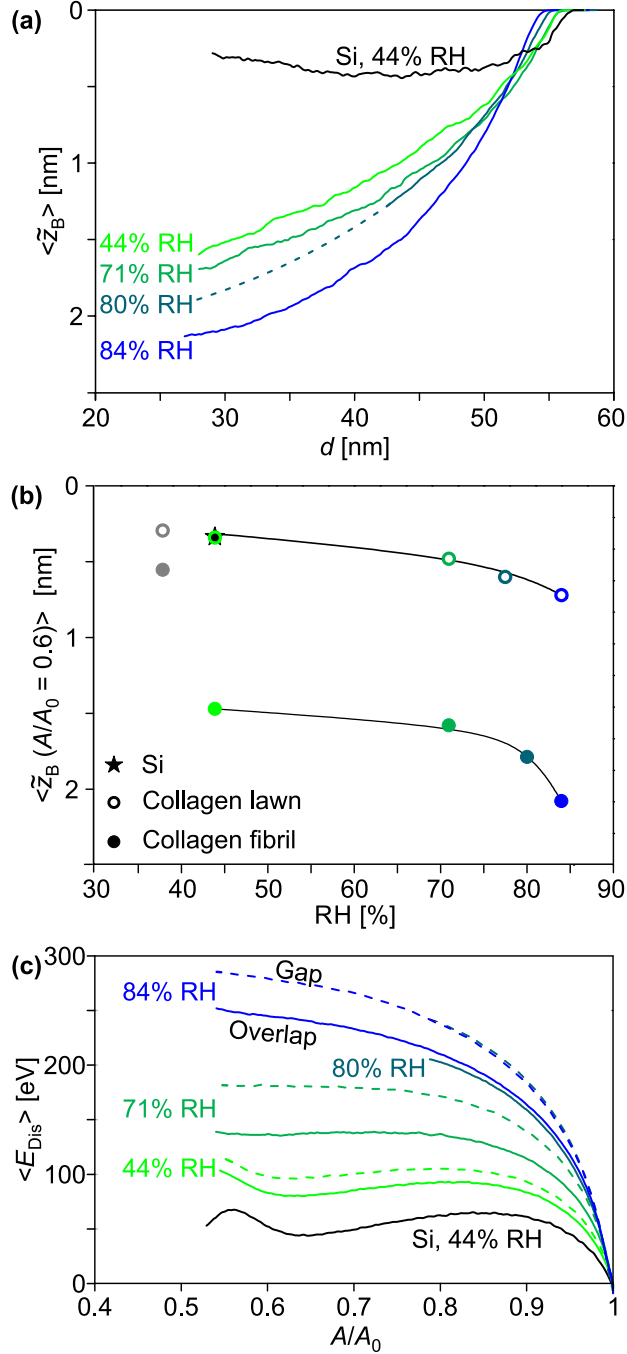

**Fig. 6.** (a) Average tip indentation $<\tilde{z}_B>$ measured on silicon (black) and on the collagen fibril (in color) plotted as a function of the mean tip–sample distance. The RH is indicated for each curve. (b) Average tip indentation $<\tilde{z}_B>$ measured at $A/A_0 = 0.6$ on silicon (star), the collagen lawn (open circles), and the collagen fibril (filled circles) as a function of the RH. The solid lines serve as a guide to the eye. The gray symbols are the data points at the end of the experiment. (c) Average dissipated energy, $<E_{Dis}>$, measured on silicon (black) and on the collagen fibril (in color) and plotted as a function of the amplitude ratio. The solid and dashed lines correspond to gap regions and overlap regions, respectively.

**Energy dissipation.** Cleveland et al.[46] and Tamayo et al.[75] showed that the phase shift in IC AFM is related to the energy dissipation during the tip–sample contact. Zitzler et al.[41] and García et al.[47] determined an expression for the dissipated energy per oscillation cycle, $E_{Dis}$, and showed that the shape of the $E_{Dis}$ curve plotted as a function of the amplitude ratio $A/A_0$ reveals information about the dissipation mechanism. Fig. 6c shows the average of the dissipated energy, $<E_{Dis}>$, measured on the Si substrate (black) and the collagen fibril's ridge (in color) as a function of the amplitude ratio $A/A_0$. The dashed lines and solid lines correspond to gap regions and overlap regions, respectively. On the Si substrate, the $<E_{Dis}>$ curve shows a plateau for $A/A_0 <$ 0.9, indicating that capillary interactions are the dominating dissipation mechanism during the tip–sample contact.[41,47] On the collagen fibril at 44% RH, the $<E_{Dis}>$ curves are qualitatively similar to the $<E_{Dis}>$ curve from measurements on the Si substrate, indicating that the dominant dissipation mechanism is also capillary interaction. However, the plateau reaches a higher value, which we ascribe to a larger tip indentation on the collagen fibril than on the Si substrate. It is also likely that on the collagen fibril, a larger amount of water is involved in the dissipation process than on the Si substrate. The FD curves measured on the collagen lawn reveal that with increasing RH, there is a larger increase in $d_{on}$ on the collagen lawn than on the Si substrate (see the Subsection Collagen lawn and Si substrate). This indicates that on the collagen fibril, the water bridge forms during the tip approach at a larger tip–sample distance than on Si. With increasing RH, the dissipated energy increases gradually, accompanied by a gradual transition of the curve shape. At 84% RH, the curve is shaped like a flattened parabolic branch. A parabolic curve shape is characteristic for viscous dissipation.[47] Hence we interpret the curve shape at 84% RH to be the result of a dissipation mechanism dominated by capillary interactions which is progressively overlaid by viscous dissipation due to the increasing tip indentation into the hydrated collagen fibril.

In gap regions, the dissipated energy is always larger than in overlap regions for all RHs. In the following, we discuss the differences in the nanomechanical properties between the gap regions and the overlap regions by considering maps of the parameters obtained from the APD measurements. Fig. 7a shows maps of the phase shift $\varphi$ reconstructed from the APD curves at an amplitude ratio of $A/A_0 = 0.75$. The phase shift signal from APD curves is the same as in conventional IC-mode AFM phase images, but the reconstructed images are free from feedback loop artifacts. Furthermore, the phase shift obtained from APD curves can be quantitatively analyzed to separate the conservative and the dissipative contributions to the tip–sample interaction.[46–48] Following Schröter et al., we determined the additional effective tip–sample spring constant, $k_{TS}$, which represents the conservative contribution to the phase shift.[48] For characterizing the dissipative interaction, we determined the dissipated energy, $E_{Dis}$, according to Cleveland et al.[46] and García et al.[47] Fig. 7b and 7c show maps of $k_{TS}$ and $E_{Dis}$, respectively, reconstructed for an amplitude set point of $A/A_0 = 0.75$. The contrast of the $k_{TS}$ maps resembles that of the $\varphi$ maps. At 44% RH, the $k_{TS}$ map shows very little contrast between the collagen fibril and the collagen lawn and a small contrast between the gap regions and overlap regions. This indicates that the stiffness of gap and overlap regions is similar at 44% RH. With increasing RH, the average $<k_{TS}>$ value on the fibril decreases, whereas on the collagen lawn, $<k_{TS}>$ has the same value at 84% RH as it does at 44% RH. The decrease in $<k_{TS}>$ is in line with the decrease in the fibril's elastic modulus (Fig. 3c) and the increase in the tip indentation $<\tilde{z}_B>$ (Fig. 6b). Moreover, the D-band contrast increases with increasing RH, with higher $k_{TS}$ values in overlap regions compared to gap regions. This indicates that overlap regions are stiffer than gap regions when the fibril is in a hydrated state, which is consistent with earlier findings.[21,22,32,33] Compared to the $\varphi$ maps, the maps of the dissipated energy, $E_{Dis}$ (Fig. 7c), show an inverted contrast. At 44% RH, the fibril is dark and only distinguishable from the lawn due to a brighter, ladder-like structure. The bright areas correspond to the gap regions, where more energy is dissipated compared to overlap regions. The same applies for the three-phase contact lines (the fibril's upper and lower edges). The dissipated energy on the collagen fibril as well as the D-band contrast increase with increasing RH.

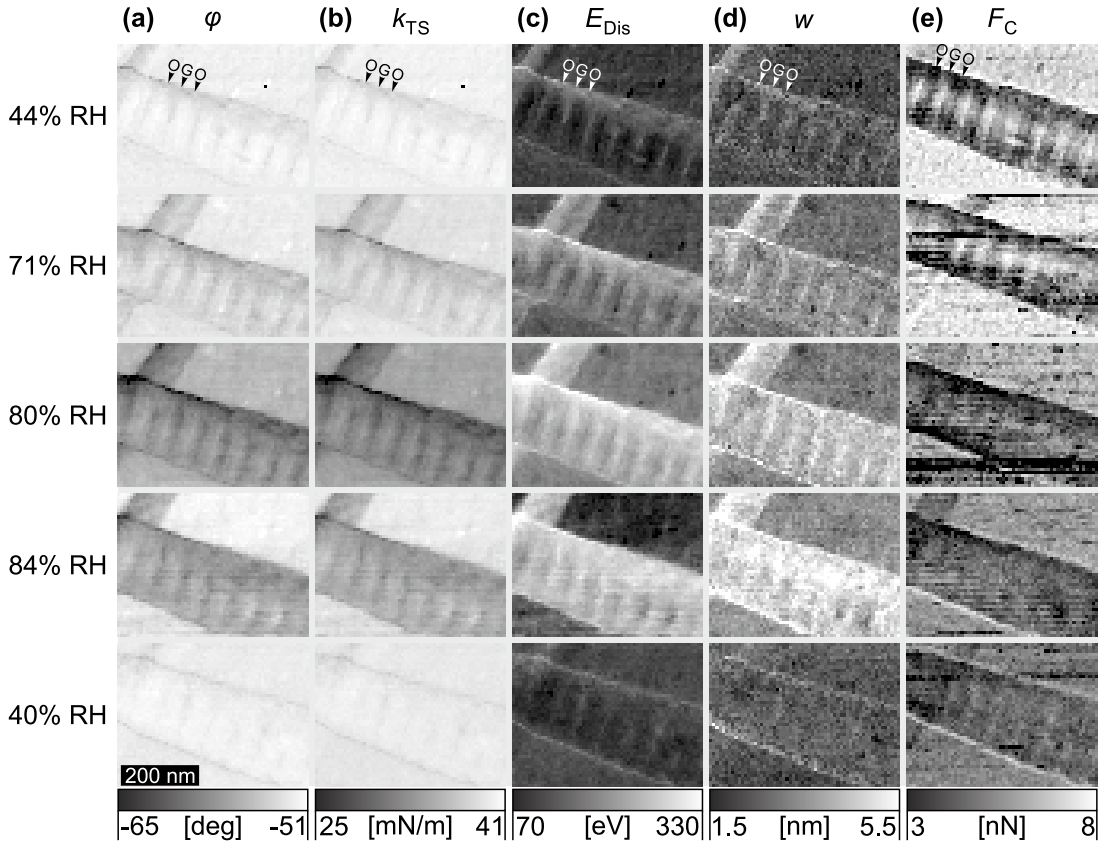

**Fig. 7.** Maps of (a) the phase, $\varphi$, (b) the conservative contribution to the tip–sample interaction, $k_{TS}$, (c) the dissipated energy, $E_{Dis}$, and (d) the attractive regime's width, $w$, obtained from APD measurements at $A/A_0 = 0.75$. (d) Maps of the maximum attractive force during approach, $F_C$, extracted from force-distance curves. The maps were measured in the area marked in Fig. 1b. For each row, the RH is indicated on the left. Gap regions and overlap regions are labelled with G and O, respectively.

Fig. 7d shows maps of the attractive regime's width $w = z_0 - z_B$, which increases with RH. In the gap (overlap) regions, it increases from 2.6 nm (2.2 nm) at 44% RH to 5.0 nm (4.0 nm) at 84% RH. Fig. 7e shows maps of the capillary force, $F_C$, obtained from FD curves. The maps are the same as in Fig. 4a and are shown in Fig. 7e for comparison. We showed that capillary interaction is the dominating dissipation mechanism at moderate RH (43% and 71% RH). Comparing the $F_C$ maps from FD measurements with the $E_{Dis}$ maps obtained from APD measurements for 43% and 71% RH shows that larger capillary forces correlate with higher dissipated energies. Hence we conclude that for RH ≤ 71%, the D-band contrast visible in the $E_{Dis}$ maps results from the capillary interaction and not from different mechanical properties of gap regions and overlap regions. The D-band structure and the three-phase contact lines visible in the $F_C$ maps are due to the fibril's local surface curvature (see Fig. 5). At 80% and 84% RH, the $E_{Dis}$ map still shows a D-band pattern, whereas the $F_C$ map does not. Therefore, at higher RH, we ascribe the larger energy dissipation in gap regions compared to overlap regions to a difference in the viscoelastic properties within the 2-nm-thick surface layer.

At the end of the experiment at 40% RH (bottom row), both the $k_{TS}$ and the $E_{Dis}$ maps show the same contrast as shown at the beginning of the experiment at 44% RH (top row). Also the average values are restored. Initially, at 44% RH, $<k_{TS}>$ = 39.7 ± 0.5 mN/m and $<E_{Dis}>$ = 100 ± 15 eV; at the end of the experiment, at 40% RH, $<k_{TS}>$ = 39.2 ± 0.3 mN/m and $<E_{Dis}>$ = 102 ± 10 eV. This shows that the hydration-related changes of the fibril's nanomechanical properties were fully reversible.

**Collagen lawn and Si substrate.**

FD and APD curves measured on the Si substrate and the collagen lawn give insight into the mechanical properties of the collagen lawn and how the tip–sample interaction is affected by the RH. In particular, we obtain information about the water film adsorbed on the collagen

Fig. 8a shows individual FD approach curves measured on a Si substrate at different RHs. With increasing RH, $d_{On}$ increases from 3 nm to 3.5 nm. Taking into account that water layers adsorb on the Si substrate and the tip apex, the increase in $d_{On}$ can be partly ascribed to an increase in the water films' thicknesses. According to Beaglehole *et al.*, the water layer adsorbed on a flat silicon oxide surface has a thickness of 0.2 nm at 44% RH and 0.4 nm at 84% RH.[37] These values correspond to the thickness of the layer of liquid water reported by Asay *et al.*[38] Also, these values are much smaller than $d_{On}$, showing that the formation of the water bridge occurs before the geometric contact of the water layers on the tip and the sample.[41,43] It is likely that the distance at which the water bridge forms increases with increasing RH due to the increase in the Kelvin radius.[76] Fig. 8b shows individual FD approach curves measured on the collagen lawn at different RHs (in color) and also an approach curve measured with the same tip on a Si substrate (black). At 44% RH, the distance $d_{On}$ measured on the collagen lawn is comparable to the value measured on the Si substrate. With increasing RH, the $d_{On}$ measured on the collagen lawn increases from 2 to 4 nm, which is four times the increase observed on the Si substrate.

According to the contact model described above (Fig. 1), we conclude that the thickness of the water film adsorbed on the collagen lawn increases more than that on the Si substrate due to the collagen lawn's hydrophilic nature. Note that the data shown in Fig. 8b were measured with a different tip than those shown in Fig. 8a. Hence the absolute $d_{On}$ values may differ due to different tip geometries.

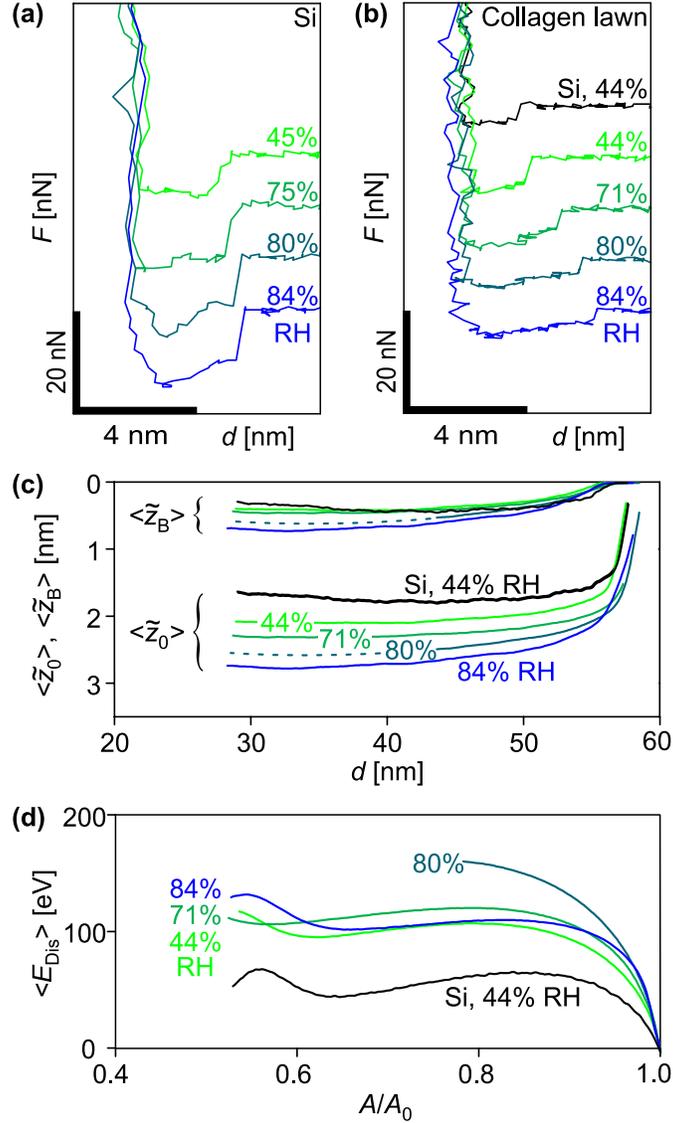

**Fig. 8.** Effects of adsorbed water layers. (a) FD approach curves measured on silicon. The curves are stacked vertically for clarity. The RH is indicated for each curve. Note that the data shown in (a) were measured with a different tip than the data shown in (b-d). (b) FD approach curves measured on silicon (black) and on the collagen lawn (in color). The curves are stacked vertically for clarity. (c) Average tip indentation, $<\tilde{z}_0>$ and $<\tilde{z}_B>$ plotted as a function of the tip–sample distance $d$, and (d) average dissipated energy, $<E_{Dis}>$, measured on the collagen lawn (in color) and on silicon (black) plotted as a function of the amplitude ratio $A/A_0$.

The difference $w = z_0 - z_B$ is the attractive regime's width. On the collagen lawn, the average $<w>$ increases from 2.2 nm at 44% RH to 3.2 nm at 84% RH. These values are comparable to the distance $d_{On}$ in the FD approach curves (Fig. 8b), which is a measure of the water bridge's length. The increase in $<w>$ and $d_{On}$ with increasing RH indicates an increase in the thickness of the water layer adsorbed on the collagen fibril, the collagen lawn, and the AFM tip. However, we point out that $<w>$ and $d_{On}$ cannot be identified with the water layer thickness, since on the Si substrate, $<w>$ and $d_{On}$ are about ten times larger than the thickness of the liquid water layers.[37,38]

Fig. 8c shows the average indentation $<\tilde{z}_0>$ and $<\tilde{z}_B>$ determined from APD curves measured on the Si substrate (black) and the collagen lawn at different RHs (in color) as a function of the tip–sample distance. The average tip indentation with respect to the collagen lawn's unperturbed surface, $<\tilde{z}_B>$, increases from 0.4 nm at 44% RH to 0.7 nm at 84% RH. Up to 71% RH, the values are the same as those measured on the Si substrate. The tip indentation $\tilde{z}_0$ with respect to the position $z_0$ is larger than $\tilde{z}_B$ and increases (on average) from 2.1 nm at 44% RH to 2.8 nm at 84% RH (measured at $A/A_0 = 0.6$).

## Conclusions

With AFM-based force spectroscopy, we studied the effect of hydration on the viscoelastic properties of a reconstituted type I collagen fibril in air with controlled relative humidity (RH). We explored the tip–sample interaction as a function of tip–sample distance and tip indentation. This allows us to separate the contributions of the sample's viscoelastic response and the capillary interaction to the tip–sample interaction. The main findings can be summarized as follows.

**Capillary interaction**

During the tip approach and retraction, a stable water bridge forms between the tip apex and the sample. With increasing RH, the tip–sample distance where the water bridge initially forms increases from 2 to 4 nm. The water bridge causes a capillary force, $F_C$, which can be measured using a stiff cantilever that avoids a snap-to-contact. The maps of $F_C$ show a D-band contrast as a result of the interplay between the collagen fibril's height undulation and the tip apex geometry. In IC-mode AFM and APD measurements, the capillary interaction is the major energy dissipation mechanism during the tip–sample contact and the main origin of the phase contrast.

**Collagen fibril's top 2 nm**

During APD measurements, the AFM tip indents up to 2 nm into the collagen fibril at 84% RH. Within these 2 nm, the hydrated collagen fibril is stiffer in overlap regions than in gap regions, which indicates a higher amount of free water in the gap regions compared to overlap regions.[22]

**Collagen fibril's bulk**

Using FD measurements, we determined the fibril's elastic modulus as a function of the RH. Within an indentation range of 5 nm, the average elastic modulus decreases from 2.63 GPa at 44% RH to 0.149 GPa at 84% RH. The large elastic modulus, the force hysteresis in the FD measurements, and the small tip indentation $z_B$ of only 2 nm at 84% show that the hydrated collagen fibril is a viscoelastic solid.

Our work demonstrates a versatile and robust methodology for studying the nanomechanical surface properties of more complex collagenous materials, such as native collagen fibrils isolated from different types of connective tissues, the tissues themselves, as well as collagen-based artificial biomaterials. From a more general perspective, the interface between condensed matter and water vapor is important for many processes in nature. We therefore expect that the methodology demonstrated in this work will be also useful for studying the influence of water vapor on the nanomechanical surface properties of other hydrated specimens, for example, hydrogels, living cells, and biological tissues.

## Acknowledgements

We thank A. Taubenberger and D. J. Müller for sharing the details of their collagen sample preparation procedure, E.-C. Spitzner for his contributions during the initial stage of the project, C. Riesch for help with the data analysis, D. Voigt and M. Dehnert for discussions, and S. McGee for proofreading the manuscript. The acquisition of the AFM was funded by the Volkswagen Foundation and the Deutsche Forschungsgemeinschaft.

TOC Entry

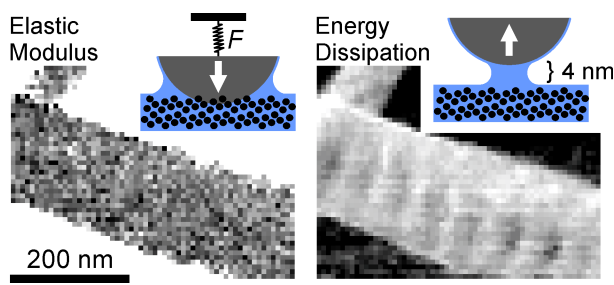

A versatile methodology for quantitative AFM imaging of hydrated specimens in humid air is presented.